\documentclass[12pt]{article}
\usepackage{graphicx}

\begin{document}

\title{\textbf{A purely reflective\\ large wide-field telescope}}

\author{V.Yu. Terebizh\thanks{98409 Nauchny, Crimea, Ukraine;
 \,E-mail:\, \textsf{vterebizh@yandex.ru}}\\
 \small{\textit{Sternberg Astronomical Institute, Moscow, Russia}}}

\date{\footnotesize{}}

\maketitle

\begin{quote}
\small{\textbf{Abstract}~--- Two versions of a fast, purely reflective
Paul-Baker type telescope are discussed, each with an 8.4-m aperture, $3^\circ$
diameter flat field and $f$/1.25 focal ratio.

The first version is based on a common, \textit{even asphere} type of surface
with zero conic constant. The primary and tertiary mirrors are 6th order
aspheres, while the secondary mirror is an 8th order asphere (referred to here
for brevity, as the 6/8/6 configuration). The $D_{80}$ diameter of a star image
varies from $0''.18$ on the optical axis up to $0''.27$ at the edge of the
field ($9.3-13.5\,\mu$m).

The second version of the telescope is based on a \textit{polysag} surface type
which uses a polynomial expansion in the sag $z$,
$$
  r^2 = 2R_0z - (1+b)z^2 + a_3 z^3 + a_4 z^4 + \ldots + a_N z^N,
$$
instead of the common form of an aspheric surface. This approach results in
somewhat better images, with $D_{80}$ ranging from $0''.16$ to $0''.23$, using
a lower-order 3/4/3 combination of powers for the mirror surfaces. An
additional example with 3.5-m aperture, $3^\circ.5$ diameter flat field, and
$f$/1.25 focal ratio featuring near-diffraction-limited image quality is also
presented.

\medskip

\textit{Key words}: General optics, telescopes
 }
\end{quote}

\section*{Introduction}

Widening the field of view of large telescopes has been an active research
topic during last few years. The special attention was given to further
development of the Mersenne~[1636] system in a direction specified by
Schmidt~[1930], resulting in a three-mirror telescope system. The important
milestones on this path were the works of Paul~[1935], Baker~[1969],
Willstrop~[1984], and Angel et. al.~[2000].

Curved focal surfaces, which are frequently encountered in wide-field designs,
remain undesirable for use with modern detectors (Ackermann, McGraw, and Zimmer
2006). As is well known, there are no practical flat-field configurations for a
Cassegrain system. The use of a lens corrector in the exit pupil of a Gregory
system enables to reach the $\sim3^\circ$ field, however it remains slightly
curved (Terebizh~2006). An excellent aberration-free solution by Korsch~[1972,
1977] provides the flat-field three-mirror designs in a frame of theory of
3rd-order aberrations, i.e., for the relatively slow systems. It was shown by
Baker~[1969] that a flat field could be attained with a fast three-mirror
telescope proposed by Paul~[1935] (see the general discussion in the book by
Schroeder~[2000], \S6.4).

Our goal was to find a purely reflecting fast three-mirror system with a flat
field not less than $3^\circ$ in diameter using only low-order aspheres. For
convenience, we consider examples of such a telescope with the physical
parameters close to those for the Large Synoptic Survey Telescope (LSST, see
Angel et. al. 2000, Seppala 2002). Two versions of such an 8.4-m, $f$/1.25
telescope are discussed.  The first telescope is based on the common,
\textit{even asphere} type surfaces with zero conic constant; the second
version is based on a \textit{polysag} type surface, which uses a polynomial
expansion in the sag $z$, instead of the common form of an aspheric surface.

\section*{8.4-m telescope based on even aspheres}

The design is shown in Fig.~1, with performance as described in Table~1, and
the complete optical prescription given in Table~A1 of the Appendix. The
prescription follows the notation used by the ZEMAX\footnote{ZEMAX Development
Corporation, U.S.A.} optical design program. Fig.~2 presents the corresponding
spot diagrams.

As one can see from Table~A1, all three mirrors are even aspheres with zero
conic constants, i.e., the slightly deformed spheres. Unlike the LSST, where
the mirrors are primarily non-spherical conic sections with the addition of
small polynomial corrections up to the 10th order, the telescope described here
uses the polynomial terms comparable with the "seed" spheres. As a consequence,
the effective radius of curvature at the vertex of each surface is noticeably
different from that of the sphere as seen in Table~A1. In this sense, our
design is closer to the Willstrop~[1984] design with only polynomial
representation of the surfaces profiles.

The key point is that the optimal choice both the conic constants and
polynomial coefficients leads to another form of the same basic design. Indeed,
we can achieve nearly the same image quality as shown in Fig.~2 for a design
with non-zero conic constants (of the order of~1) and a new set of the
polynomial coefficients. \textit{A whole set of parameters simply adjust the
theoretically best surface profiles given flat image}. The existence of the
objectively best profiles for surfaces in a three-mirror telescope with a flat
focal surface appears to be natural in the context of the Schwarzschild~[1905]
approach to aplanatic systems and generalization of that approach to any
two-mirror aplanats (Terebizh~2005). We will address this topic later in more
detail.

\begin{figure}[t]   
   \centering
   \includegraphics[width=0.65\textwidth]{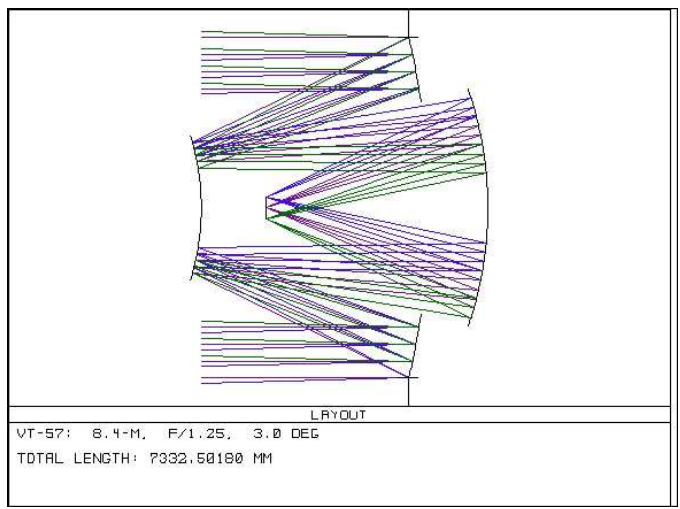}
   \caption*{First version of a 8.4-m telescope.}
\end{figure}

\begin{center}   
\small{
\begin{tabular}{|l|c|}
 \multicolumn{2}{l}{\textit{Table 1. First version of the 8.4-m telescope}}\\[3pt]
 \hline
 &\\
 \multicolumn{1}{|c|}{Parameter} & Value\\
 &\\
 \hline
 &\\
 Entrance pupil diameter                & 8400 mm\\
 Effective diameter                     & \\
 \quad center of field -- edge          & 6560 -- 6470 mm\\
 Effective focal length                 & 10500 mm \\
 Effective $f$-number                   & 1.25 \\
 Scale in the focal plane               & 50.905 $\mu$m/arcsec \\
 Angular field of view                  & $3^\circ.0$ \\
 Linear field of view                   & 550.4 mm \\
 Image RMS-diameter                     & \\
 \quad center of field -- edge          & $0''.13-0''.19$, 6.7 -- 9.8 $\mu$m \\
 Image $D_{80}$ diameter                & \\
 \quad center of field -- edge          & $0''.18-0''.27$, 9.3 -- 13.5 $\mu$m \\
 Maximum distortion                     & 0.087\% \\
 Fraction of unvignetted rays           & \\
 \quad center of field -- edge          & 0.610 -- 0.593 \\
 Orders of the aspheric mirrors         & 6/8/6\\
 Length of the optical system           & 7333 mm \\
 \hline
\end{tabular}
 }
\end{center}

\begin{figure}[t]   
   \centering
   \includegraphics[width=0.65\textwidth]{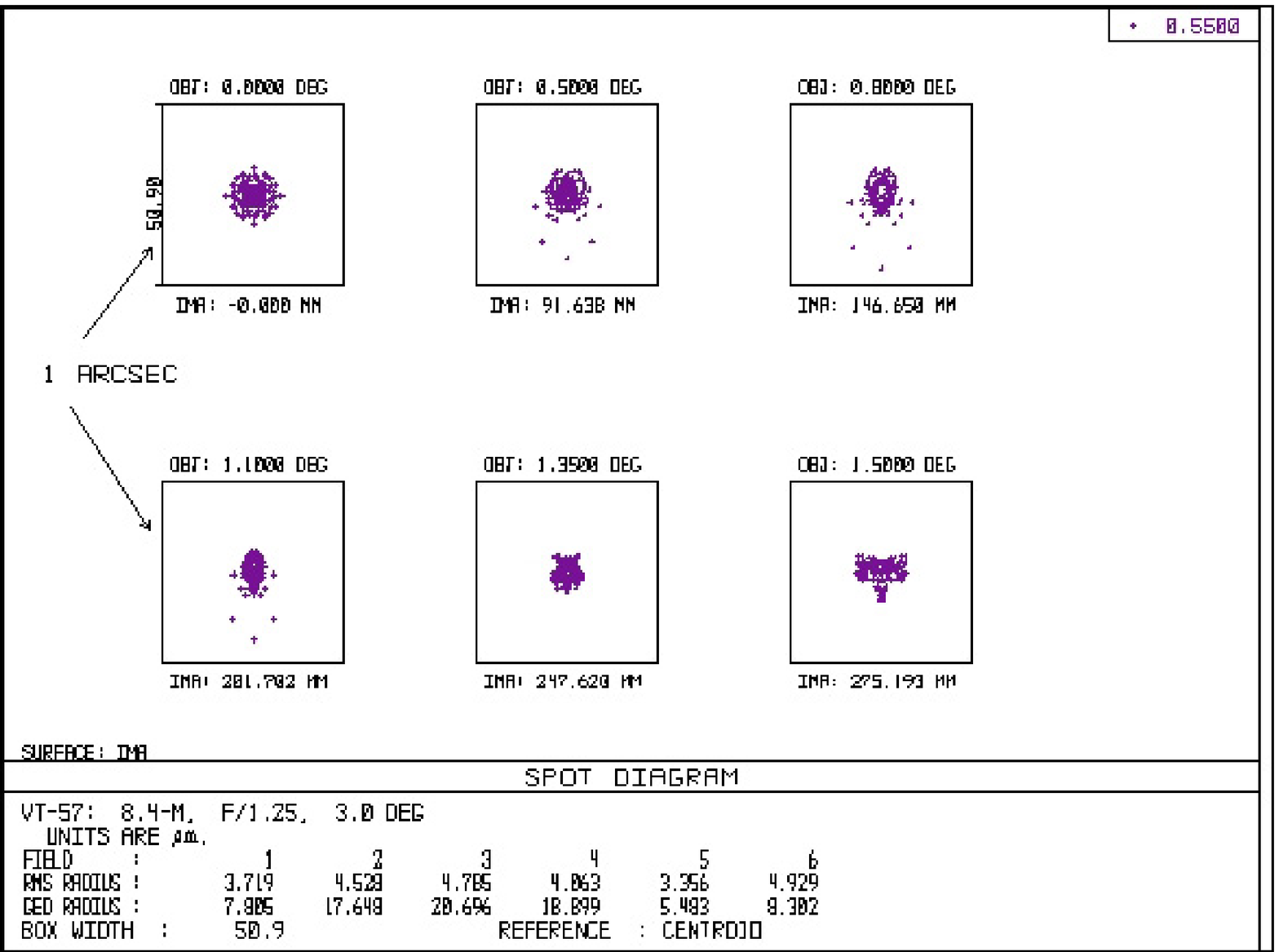}
   \caption*{Spot diagrams of the telescope shown in Fig.~1\\
    for the field angles $0,\, 0^\circ.5,\, 0^\circ.8,\, 1^\circ.1,\, 1^\circ.35$,
    and $1^\circ.5$.\\
    Wavelength is $0.55\,\mu$m, the box width is $1''$ ($50.9\,\mu$m).}
\end{figure}

\begin{center}   
\small{
\begin{tabular}{|c|l|c|c|c|}
 \multicolumn{5}{l}{\textit{Table 2. Properties of the nearest spheres}}\\[3pt]
 \hline
&&&&\\
& Mirror & $R_n$, mm & $\phi_n$ & $\delta_n$, mm \\
&&&&\\
\hline
 Present       &Primary       & -- 18563.048 & 1.105 & -- 1.728 \\
 design        &Secondary     & -- 6046.431  & 0.854 & 0.0698   \\
               &Tertiary      & -- 9097.721  & 0.771 & 0.549    \\
 \hline
               &Primary       & -- 17998.234 & 1.071 & -- 1.670  \\
 LSST          &Secondary     & -- 6167.953  & 0.916 & -- 0.0696 \\
               &Tertiary      & -- 8411.999  & 0.774 & -- 0.214  \\
 \hline
\end{tabular}
 }
\end{center}

\bigskip

Taking into account the above discussion concerning surface profiles, it is
interesting to find their deviation from the \textit{nearest} sphere.  It is
sufficient for our purposes to choose the simple definition of the nearest
sphere, namely, the sphere that includes the surface's vertex and outward rim.
Table~2 gives the radiuses of the nearest spheres $R_n$, the corresponding
$f$-numbers $\phi_n \equiv |R_n|/(2 D_n)$, and the maximum by modulus deviation
of each surface from the nearest sphere $\delta_n \equiv z_s-z_n$ for the
design presented here and the LSST. In general, both sets of parameters are
comparable in value.

\section*{The \textit{polysag} type optical surfaces}

An optical surface which is symmetric about the $z$-axis is usually described
by a conic section equation
$$
  r^2 = 2R_0z - (1+b)z^2,
  \eqno(1)
$$
where $r = \sqrt{x^2 + y^2}$ is the radial coordinate, $R_0$ is the paraxial
radius of curvature, and the conic constant $b$ is equal to the negative of the
squared eccentricity: $b = -\varepsilon^2$. In optical ray tracing, it is
suitable to solve equation~(1) with respect to the sag $z$, so the
\textit{standard} surface form is defined by equation
$$
  z = \frac{r^2/R_0}{1+\sqrt{1-(1+b)(r/R_0)^2}}\,.
  \eqno(2)
$$
In designing fast, wide-field optical systems, the conic sections are often
insufficient tools, so a polynomial in the radial coordinate is added to the
sag representation~(2). For example, an \textit{even asphere} surface is
defined as follows:

$$
  z = \frac{r^2/R_0}{1+\sqrt{1-(1+b)(r/R_0)^2}} + \alpha_1 r^2 +
   \alpha_2 r^4 + \ldots + \alpha_N r^{2N}.
  \eqno(3)
$$
These surfaces are quite useful in practice, but some limitations become
apparent when we enlarge the system's aperture, speed and the field of view.

\begin{figure}[t]   
   \centering
   \includegraphics[width=0.65\textwidth]{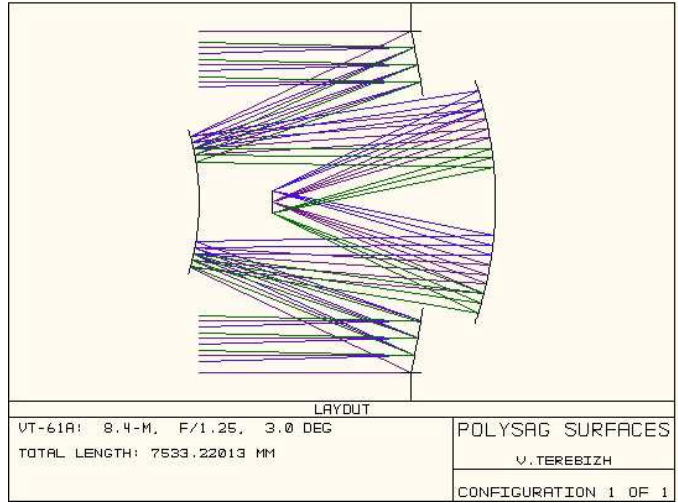}
   \caption*{Second version of a 8.4-m telescope.}
\end{figure}

The point is that a power series slowly converges to a desired function (see,
e.g., Lanczos~[1988], Ch.~7; Press et. al.~[1992], \S5.1). In optics, we seek
as proximate as possible representation of the optimal surface profile, so we
are really interested in the most quickly converging series. Meanwhile, the
convergence of a power representation~(3) is especially slow for fast systems
of large aperture, because~(3) deals with powers of the ratio $r/R_0$, which is
not particularly small near the edge of the aperture.

For these reasons, another polynomial approximation can be used to reach the
better convergence. The known for a long time expansion

$$
  r^2 = 2R_0z - (1+b)z^2 + a_3 z^3 + a_4 z^4 + \ldots + a_N z^N
  \eqno(4)
$$
can be considered as a natural generalization of equation~(1) for the conic
sections (see, e.g., Rusinov~[1973]). Here $a_3, a_4,\ldots,a_N$ are
coefficients which along with $R_0$ and $b$ define a polynomial representation
\textit{in the sag $z$, but not in the radial coordinate}~$r$. Even for very
fast surfaces, we have usually ${z \ll r}$, so the \textit{poly}nomial
expansion in the \textit{sag} (\textit{polysag}) is expected to converge more
quickly than~(3). Besides, the direct extension of equation~(1) in powers of
the sag appears sometimes to be a more logical approach than adding a series in
powers of $r$ to the solution of equation~(1) with respect to the sag.

\begin{figure}[t]   
   \centering
   \includegraphics[width=0.65\textwidth]{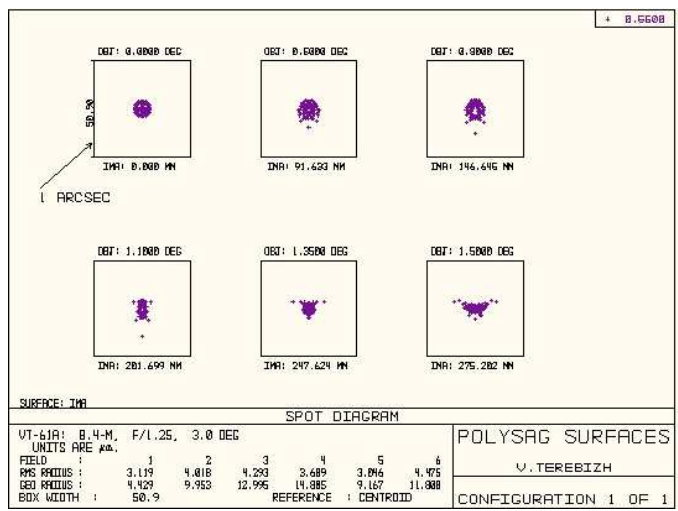}
   \caption*{Spot diagrams of the telescope shown in Fig.~3\\
    for the field angles $0,\, 0^\circ.5,\, 0^\circ.8,\, 1^\circ.1,\,
     1^\circ.35$, and $1^\circ.5$.\\
    Wavelength is $0.55\,\mu$m, the box width is $1''.0$ ($50.9\,\mu$m).}
\end{figure}

The decomposition~(4) was previously tried at low orders ${N = 3}$ or $4$.
Modern computers enable to use the polynomial expansions of any order. The only
problem is that advanced optical programs, e.g., ZEMAX, are based on the
sag~$z$ as a function of the radial coordinate~$r$, while~(4) gives the inverse
relation. This problem, however, can easily be solved resulting the additional
surface type, the \textit{polysag}\footnote{The corresponding
\textit{us\_polysag.dll} file for ZEMAX can freely be received by e-mail.}.

Examples presented below show that \textit{polysag} surfaces of relatively low
order allow one to design large and fast telescopes with high image quality.
However, this type of surface is not universal, as sometimes it is better to
treat a surface, with the aid of equation~(3). An example is the corrector
plate for a Schmidt camera; perhaps, the reason is that the spherical
aberration of a sphere is determined just by power terms in the radial
coordinate and not in the sag. Clearly, further work with the \textit{polysag}
type surface before limitations of its general application can be well
understood.

\section*{8.4-m telescope based on \textit{polysag} surfaces}

Our \textit{polysag}-based 8.4-m design has the same physical parameters as the
first example with the traditional \textit{even asphere} surfaces.  The optical
layout of the telescope is depicted in Fig.~3, with the corresponding spot
diagrams shown in Fig.~4, and its performance and parameters given in Tables~3
and~A2.

\begin{center}   
\small{
\begin{tabular}{|l|c|c|}
 \multicolumn{3}{l}{\textit{Table 3. Performance of the polysag-based
  telescopes}}\\[3pt]
 \hline
 &&\\
 \multicolumn{1}{|c|}{Parameter} & 8.4-m & 3.5-m \\
 &&\\
 \hline
 &&\\
 Entrance pupil diameter         & 8400 mm                 & 3500 mm \\
 Effective diameter              &                         & \\
  \quad center of field -- edge  & 6561 -- 6452 mm         & 2711 -- 2711 mm \\
 Effective focal length          & 10500 mm                & 4375 mm \\
 Effective $f$-number            & 1.25                    & 1.25 \\
 Scale in the focal plane        & 50.905 $\mu$m/arcsec    & 21.211 $\mu$m/arcsec\\
 Angular field of view           & $3^\circ.0$             & $3^\circ.5$\\
 Linear field of view            & 550 mm                  & 268 mm \\
 Image RMS-diameter              &                         & \\
  \quad center of field          & $0''.12$ (3.1 $\mu$m)   & $0''.21$ (4.4 $\mu$m)\\
  \quad edge                     & $0''.18$ (9.0 $\mu$m)   & $0''.30$ (6.3 $\mu$m)\\
 Image $D_{80}$ diameter         &                         & \\
  \quad center of field          & $0''.16$ (8.3 $\mu$m)   & $0''.29$ (6.1 $\mu$m)\\
  \quad edge                     & $0''.23$ (11.6 $\mu$m)  & $0''.41$ (8.7 $\mu$m)\\
 Maximum distortion              & 0.09\%                  & 0.12\%\\
 Fraction of unvignetted rays    &                         & \\
 \quad center of field -- edge   & 0.61 -- 0.59            & 0.60 -- 0.60 \\
 Orders of the aspheric mirrors  & 3/4/3                   & 3/4/3 \\
 Length of the optical system    & 7533 mm                 & 3485 mm \\
 \hline
\end{tabular}
 }
\end{center}

\bigskip

The use of the \textit{polysag}-type mirrors enables the design to achieve even
the better image quality on a flat field with a 3/4/3 configuration of mirrors.
The image quality approximately matches atmosphere seeing with spot sizes
closely matched to the size of commonly used detector pixels. Note that
dimensions of the image spots are not far from the diameter of the central peak
in the Airy pattern (1.7~$\mu$m).

As one might expect from general properties of the Paul-Baker telescope, the
primary mirror for the design shown in Fig.~3 is close to a paraboloid, while
the secondary and tertiary mirrors are, in the first approximation, spherical.
Unlike the first design, for the \textit{polysag} type telescope specified in
Table~A2 the radiuses of curvature $R_0$ are the paraxial radiuses of the
\textit{surfaces}, but do not describe only their spherical components (compare
$R_0$ from Table~A2 with the $R_n$ values for the first design given in
Table~2).

\begin{figure}[t]   
   \centering
   \includegraphics[width=0.65\textwidth]{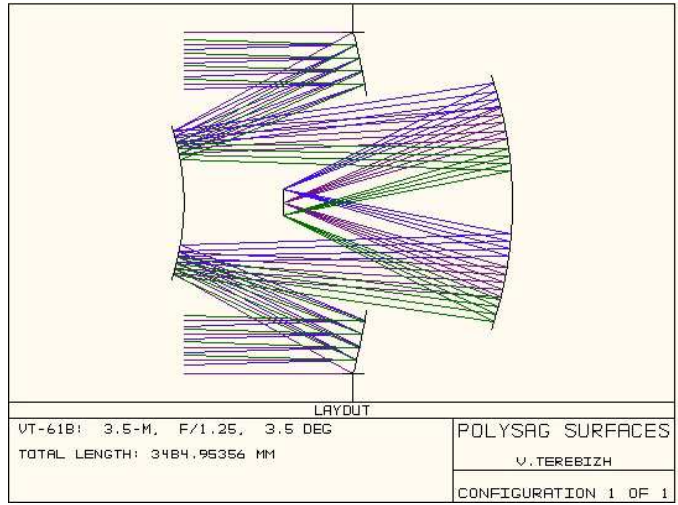}
   \caption*{All-reflective 3.5-m telescope\\ with a flat field of view of
    $3^\circ.5$ in diameter.}
\end{figure}

\section*{3.5-m telescope based on \textit{polysag} surfaces}

A wider field of view can be attained by applying the \textit{polysag} surfaces
of higher order, or by scaling down the 8.4-m designs discussed above. As a
third example, using the \textit{polysag} type surfaces, we consider 3.5-m
telescope with a flat $3^\circ.5$ field. This system is presented in Figs.~5
and~6; with performance values given in Table~3. As one can see, the 3.5-m
design has nearly diffraction-limited image quality.

\begin{figure}[t]   
   \centering
   \includegraphics[width=0.65\textwidth]{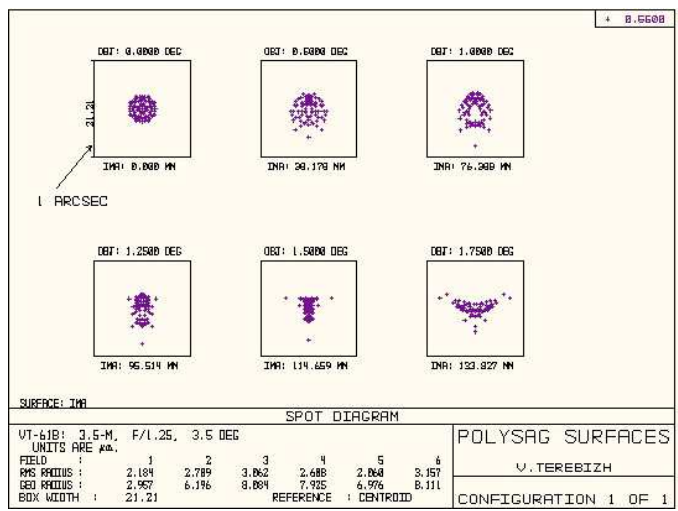}
   \caption*{Spot diagrams of the telescope shown in Fig.~5\\
    for the field angles $0,\, 0^\circ.5,\, 1^\circ.0,\, 1^\circ.25,\,
    1^\circ.5$, and $1^\circ.75$.\\
    Wavelength is $0.55\,\mu$m, the box width is $1''.0$ ($21.2\,\mu$m).}
\end{figure}

\section*{Concluding remarks}

In practice, each of the telescopes considered above should be supplied with
additional optical elements such as a filter and detector window. This,
however, is not a simple problem at $f/\# \simeq 1.25$, because of longitudinal
chromatic aberration (see Schroeder~[2000] for the discussion and examples).
Although a nearly afocal pair of filter and window lenses could be introduced
with an acceptable loss of image quality, the design loses some advantages of
the initial all-reflective telescope.  Use of thin flat plates as the filter
and window appears to be a preferable solution. For example, placing a plate
5~mm thick approximately 10~mm from the detector in the 8.4-m design results in
image deterioration of roughly $20\%$. According to M.~Ackermann~(private
communication), this approach is currently being studied at the Sandia National
Laboratories.

For some kinds of observations, image distortion may be important. We did not
specifically restrict distortion in the proposed designs. Its value of 0.087\%
for the first design can be reduced at least a factor of two with negligible
loss of the image quality.

We choose the $3^\circ.0$ field mainly to facilitate comparison with other
systems; perhaps, one can reach the wider field even for a 8.4-m aperture. As
mentioned previously, scaling down the designs discussed above is a useful way
for creating attractive telescopes with a wider field and simpler surfaces.
Indeed, it is much easier to restrict aberrations and obscuration for smaller
systems. Also, necessary components such as filters and windows can be added
while still maintaining high image quality with the addition of a lens field
corrector. This is allowable for the ground-based telescopes, but for the space
systems the inherent advantage of the purely reflective optics cannot be
matched.

\bigskip

The author is grateful to M.R.~Ackermann and V.V.~Biryukov for useful
discussions.

\newpage

\section*{Appendix:\\ The complete description of the designs}

\bigskip

\begin{center}   
\small{
\begin{tabular}{|c|l|c|c|c|c|}
 \multicolumn{6}{l}{\textit{Table A1. First version of the 8.4-m
 telescope$^{a}$}}\\[3pt]
\hline
 &&&&&\\
 Number  &         &Curvature &Thickness &      &Light     \\
 of the  &Comments &radius    &(mm)      &Glass &diameter  \\
 surface &         &(mm)      &          &      &(mm)      \\
 &&&&&\\
\hline
 &&&&&\\
 1  &Shield$^{b}$           &$\infty$    &5121.208    &---     &3540.0 \\
 2  &Aperture stop$^{c}$    &$\infty$    &481.379     &---     &8400.0 \\
 3  &Primary$^{d}$          &$-20556.85$ &$-5602.588$ &Mirror  &8400.0 \\
 4  &Secondary$^{e}$        &$-26012.21$ &$5602.588$  &Mirror  &3538.1 \\
 5  &Beam on primary$^{f}$  &$-20556.85$ &$1465.335$  &---     &5500.0 \\
 6  &Tertiary$^{g}$         &$-8619.046$ &$-5477.544$ &Mirror  &5900.0 \\
 7  &Image$^{h}$            &$\infty$    &            &        &550.4  \\
\hline
\end{tabular}
 }
\end{center}
 {\footnotesize
 $^{a)}$ All conic constants are equal to zero.\\
 $^{b)}$ Standard surface. Circular obscuration between radiuses 0.0
  and 1770.0~mm.\\
 $^{c)}$ Standard surface.\\
 $^{d)}$ Even asphere surface with $\alpha_1 = -3.001299e-006$,
 $\alpha_2 = 1.657118e-014$, and $\alpha_3 = 6.364267e-024$. Circular
  aperture between radiuses 2620.0 and 4200.0~mm.\\
 $^{e)}$ Even asphere surface with $\alpha_1 = -6.349123e-005$,
 $\alpha_2 = -5.017399e-013$, $\alpha_3 = -2.061083e-020$, and
 $\alpha_4 = -1.168727e-027$. Circular aperture between radiuses 0.0 and
  1770.0~mm.\\
 $^{f)}$ Even asphere surface picked up from the primary mirror. Circular
 aperture between radiuses 0.0 and 2620.0~mm.\\
 $^{g)}$ Even asphere surface with $\alpha_1 = 3.283586e-006$,
 $\alpha_2 = 7.158976e-015$, and $\alpha_3 = -1.745620e-022$. Circular
  aperture between radiuses 0.0 and 2950.0~mm.\\
 $^{h}$ Standard flat surface.
 }

\newpage

\begin{center}   
 \small{
\begin{tabular}{|c|l|c|c|c|c|c|}
 \multicolumn{7}{l}{\textit{Table A2. Second version of the 8.4-m
 telescope}}\\[3pt]
\hline
 &&&&&&\\
 Number  &         &Curvature &Thickness &      & Light    & Conic \\
 of the  &Comments &radius    &(mm)      &Glass & diameter &       \\
 surface &         &(mm)      &          &      & (mm)     &       \\
 &&&&&&\\
\hline
 &&&&&&\\
 1  &Shield$^{a}$         &$\infty$    &5196.805    &---    &3600.0 & 0 \\
 2  &Aperture stop$^{b}$  &$\infty$    &476.775     &---    &8400.0 & 0 \\
 3  &Primary$^{c}$        &$-18481.55$ &$-5673.58$  &Mirror &8400.0 &$-1.083191$\\
 4  &Secondary$^{d}$      &$-6080.58$  &$5673.58$   &Mirror &3525.3 &$-0.041400$\\
 5  &Beam &&&&&\\
    & on primary$^{e}$    &$-18481.55$ &$1598.275$  &---    &5500.0 & 0 \\
 6  &Tertiary$^{f}$       &$-9165.65$  &$-5485.393$ &Mirror &5950.0 &0.146856\\
 7  &Image$^{g}$          &$\infty$    &            &       &550.4  & 0 \\
\hline
\end{tabular}
 }
\end{center}
 {\footnotesize
 $^{a)}$ Standard surface. Circular obscuration between radiuses 0.0
  and 1800.0~mm.\\
 $^{b)}$ Standard surface.\\
 $^{c)}$ Polysag surface with $a_3 = 2.175935e-005$. Circular
  aperture between radiuses 2620.0 and 4200.0~mm.\\
 $^{d)}$ Polysag surface with $a_3 = 3.175638e-004$, $a_4 = -9.861486e-008$.
  Circular aperture between radiuses 0.0 and 1770.0~mm.\\
 $^{e)}$ Polysag surface picked up from the primary mirror.
  Circular aperture between radiuses 0.0 and 2620.0~mm.\\
 $^{f)}$ Polysag surface with $a_3 = 1.265684e-005$.
  Circular aperture between radiuses 0.0 and 2975.0~mm.\\
 $^{g}$ Standard flat surface.
 }

\end{document}